\documentclass[12pt]{iopart}

\usepackage{iopams}
\usepackage{graphicx}
\usepackage{epstopdf}  
\usepackage{amsfonts}
\usepackage{amssymb}

\newcommand{\ptt}[1]{\frac{\partial#1}{\partial t}}
\newcommand{\vvec}{\mathbf{v}}
\newcommand{\Bvec}{\mathbf{B}}

\newcommand{\Jvec}{\mathbf{J}}
\newcommand{\Avec}{\mathbf{A}}

\begin{document}

\title{Turbulence analysis of an experimental flux rope plasma}

\author{D.A. Schaffner$^{1}$, V.S. Lukin$^{2}$, A. Wan$^{1}$ and M.R. Brown$^{1}$}

\address{$^{1}$ Swarthmore College, Swarthmore, PA, USA}
\address{$^{2}$ Space Science Division, Naval Research Laboratory, Washington, DC, USA}
\begin{abstract}
We have previously generated elongated Taylor double-helix flux rope plasmas in the SSX MHD wind tunnel.  These plasmas are remarkable in their rapid relaxation (about one Alfv\'en time) and their description by simple analytical Taylor force-free theory despite their high plasma $\beta$ and high internal flow speeds. We report on the turbulent features observed in these plasmas including frequency spectra, autocorrelation function, and probability distribution functions of increments.  We discuss here the possibility that the turbulence facilitating access to the final state supports coherent structures and intermittency revealed by non-Gaussian signatures in the statistics.  Comparisons to a Hall-MHD simulation of the SSX MHD wind tunnel show similarity in several statistical measures.
\end{abstract}


\maketitle

\section{Overview}

Flux ropes observed in the heliosphere have two striking properties.   First is their rapid emergence.  Whether in the magnetosphere or in the solar corona, these large scale structures emerge rapidly, often in just a few Alfv\'en crossing times of the system.  Second is their long lifetimes.  Once formed, these structures persist for long times despite being embedded in turbulent MHD plasma.  

Flux ropes have recently been observed {\it in situ} at the subsolar magnetopause~\cite{Oieroset11}. Since magnetospheric flux ropes evolve rapidly, observations of these flux ropes tend to be made at later stages of their evolution.  However, in this remarkable coordinated observation using three THEMIS spacecraft, a flux rope is caught in the process of forming, revealing properties that are fundamentally three-dimensional (3D). 

Flux ropes are also observed remotely in the solar atmosphere~\cite{Patsourakos13}.  On July 19, 2012, an eruption occurred on the solar surface producing dynamical magnetic activity resulting in a destabilized flux rope, the acceleration of a fast ($\approx 1000~km/s$) coronal mass ejection, and a long-lived solar coronal arcade.  The long-lived structure is remarkable in its nearly semi-circular geometry, and the persistence of a ``coronal rain'' from the loop tops for nearly 24 hours.  The observation was made with the Solar Dynamics Observatory's AIA instrument on the sun's lower right hand limb.   This represents the first direct evidence of a fast CME driven by the prior formation and destabilization of a coronal magnetic flux rope.


In a laboratory plasma, Gray, {\it et al.}~\cite{Gray13} recently reported on the observation of a long-lived helical flux rope called a Taylor double-helix in the SSX MHD wind tunnel.  The Taylor double-helix is the natural relaxed state of MHD plasma confined in a long, perfectly conducting cylinder~\cite{Taylor86}.  In the case of an infinite cylinder of radius $a$, the minimum energy state has a helical pitch of $ ka = 1.234$, where k is the wave number associated with the axis of the cylinder.

In the SSX experiments, a magnetized plasma gun launches a magnetized plasma plume into a long flux conserving cylinder.  The plasma rapidly relaxes to the double-helix state in about 1 Alfv\'en crossing time and subsequently decays resistively.  Gray, {\it et al.}~\cite{Gray13} postulated that the physics of selective decay was at play as the initially turbulent plasma relaxed to the double-helix state.  The selective decay hypothesis posits that the energy selectively decays relative to the magnetic helicity because the energy spectra peaks at higher wave numbers, where dissipation is higher~\cite{Matthaeus80}.  The wind tunnel's minimum energy state possesses $ka = 1.292$, which is within 5\% of the infinite cylinder's $ka = 1.234$. 

Servidio {\it et al.}~\cite{Servidio08,Servidio11} detail simulations which observe the rapid and simultaneous magnetohydrodynamic relaxation into localized patches of plasma with near alignment of $\Bvec$ and $\Jvec$. These patches of locally relaxed plasma can then negotiate with adjacent patches to reach a globally relaxed state on a longer time scale.  However, many of the characteristics of the relaxed state will be evident locally. This localized relaxation might explain the rapidity of the transition observed in the double-helix plasmas.  A fully relaxed Taylor state would be expected to have a flat lambda profile (where $\nabla\times\Bvec = \lambda\Bvec$ governs the equilibrium).  The reported lack of a flat radial lambda profile could also be a consequence of a patchy relaxation.

We consider the possibility that the MHD turbulent flow in the SSX wind tunnel contains patches of locally relaxed plasma with reconnection sites at the boundaries.  A fully relaxed flow might be expected to exhibit Gaussian statistics in its fluctuations and power law behavior for the power spectra; conversely, a flow containing coherent structures and reconnection sites should exhibit non-Gaussian statistics.  Simulations show that coherent structures appear rapidly, in less than one dynamical time.  Large numbers of reconnection sites can be identified statistically in MHD turbulence studies \cite{Servidio09,Servidio10a}.  A statistical way to find these coherent structures is to identify rapid changes in the magnetic field vector.  A useful technique is to generate a probability distribution function (PDF) of vector increments \cite{Greco08,Greco09}.

Greco, {\it et al}~\cite{Greco08} identified intermittent structures in MHD turbulence simulations using statistical techniques, then connected the structures with regions of high current density.  Using data from a high resolution 3D MHD simulation, PDFs of magnetic field vector increments $\Delta\Bvec = |\Bvec(s + \Delta s) - \Bvec(s)|$ were constructed for two scale separations $\Delta s$; one much smaller than the correlation scale $\lambda_C$ and another $\approx 2\lambda_C$.  It was found that the PDF of $\Delta\Bvec$ at the larger increment was close to Gaussian indicating that increments larger than the correlation scale are normally distributed.  However, the PDF at the smaller increment had substantial tails in the distribution and furthermore, had the same distribution as the PDF of a component of the current density, indicating that the non-Gaussian statistics were correlated with intense current sheets.  The identification of discontinuities using this spatial interval method was also compared with coherent structures identified using intermittency statistics and found the techniques to be similar~\cite{Greco08}.

In a follow-up investigation, Greco, {\it et al}~\cite{Greco09} studied the statistics of ACE solar wind data as well as 2D and 3D simulation data using the same techniques.  Time series were analyzed for the solar wind data while spatial separations were analyzed from the simulations.  First, it was found that the ACE solar wind data had nearly identical increment PDFs to the 2D simulation data.  Second, specific features in the 2D simulation were shown to be correlated to departures from Gaussian statistics in the PDF: (a) a narrow inner peak is super-Gaussian and corresponds to small fluctuations in the lanes between magnetic islands; (b) an intermediate range is sub-Gaussian and corresponds to fluctuations in current cores inside magnetic flux tubes; (c) at several standard deviations, there are super-Gaussian wings corresponding to coherent small-scale current-sheet-like structures that form the sharp boundaries between the magnetic flux tubes.

Non-Gaussian statistics and characteristic coherent structures are initiated almost identically in weakly dissipative and ideal systems~\cite{Wan09}. Therefore we postulate that the origins of coherence and intermittency are essentially ideal, with dissipation acting only to limit growth of the smallest scale structures.  We expect that the modest value of SSX MHD wind tunnel Lundquist number ($S \sim 1000$) should not impact the emergence of coherent structures.

In this manuscript, we focus on the post-selective decay phase of the double-helix, in particular, processes that rapidly evolve the state such as patchiness and the evolution of coherent structures. First, we report on the turbulent temporal power spectrum for various aspects of these experimental flux ropes and fit to power-laws in order to compare to similar measurements in both the solar wind and MHD simulations. Then, we present MHD turbulence statistics that suggest the emergence of non-Gaussian structures. These experimental turbulent features are compared to synthetic data extracted from simulations of an MHD wind tunnel designed to have a geometry and plasma parameters matched to the SSX experiment. The simulations use both resistive-MHD and Hall-MHD numerical models within the HiFi framework~\cite{Lukin11}.

\section{Experiment}

\begin{figure}[!htbp]
\centerline{
\includegraphics[width=8.5cm]{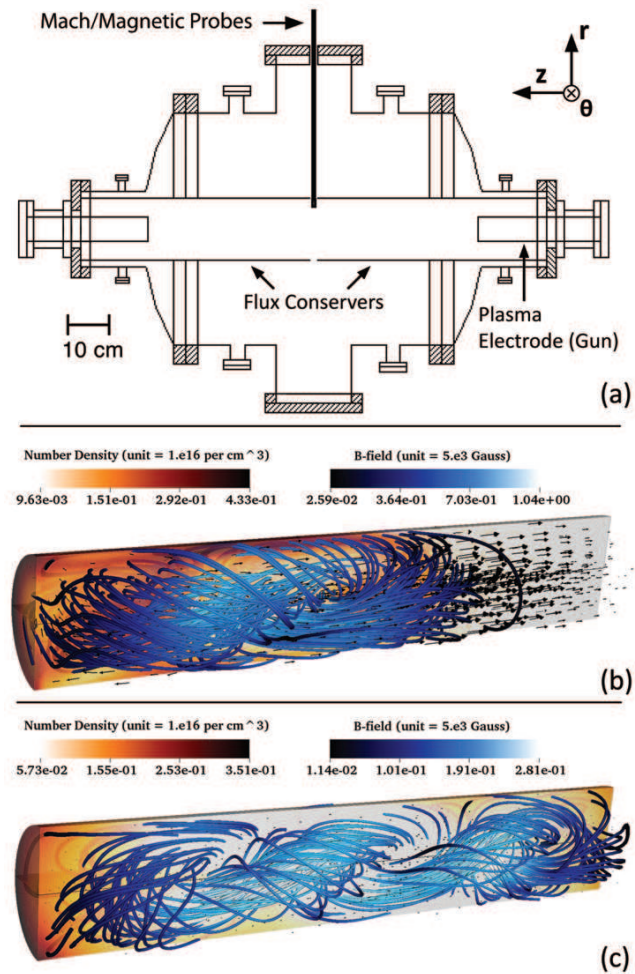}}
\caption{\label{fig:SSXdiagram} (a) SSX diagram with the two plasma gun configuration. The plasma is confined to the cylindrical flux conserver in the manner of the simulated images in (b) and (c). Magnetic and Mach probes are indicated at the midplane while interferometer density is measured $21.5~cm$ off the midplane. (b) \& (c) B-field structure, plasma number density distribution, and ion velocity vectors from the HiFi Hall-MHD simulation, with the plasma source located on the left.  An early time, before selective decay, is shown in (b) and a later time after selective decay is shown in (c). [see online version for the full simulation movie corresponding to these stills in (b) \& (c).]}
\end{figure}

The flux ropes under investigation are formed in a wind-tunnel configuration of the Swarthmore Spheromak Experiment. A copper cylindrical flux conserver serves as the tunnel capped by two plasma gun electrodes whose extent limits the length of the tunnel to $86~cm$ as can be seen in Figure~\ref{fig:SSXdiagram}. The radius, R,  of the cylinder is $7.75~cm$ making the aspect ratio of the configuration, L/R = 11.1. Though slightly shorter than the tunnel reported in previous work~\cite{Gray13}, the aspect ratio is considered still large enough for comparison to an infinite cylinder in Taylor relaxation theory. The plasma itself is formed by the discharge of a $1~mF$, $3.5~kV$ capacitor across a few centimeters wide gap between the tungsten-coated gun inner electrode and the outer wall into a puffed volume of hydrogen gas. After ionization, currents of over $80~kA$ across the gap push plasma into the main section of the flux conserver through $\Jvec\times\Bvec$ forces. Magnetic coils coaxial to the gun electrode and flux conserver contribute the stuffing flux which allows for the formation of a spheromak at the gun edge. Given the high aspect ratio, the spheromak tilts, eventually forming a twisted double-helix Taylor state; this sequence has been shown to occur in a very short time span~\cite{Gray13}. For this discharge configuration, the amount of current needed to push spheromaks off of the gun and into the main chamber is maintained for approximately $30~\mu s$ ($\sim 20-50~\mu s$ from initial discharge). The discharge likely begins with the formation of a spheromak structure into the chamber vacuum, which may or may not completely break-off from the gun source before more plasma is introduced behind it. Thus these shots generally consist of a sustained injection of structured plasma rather than a single well-defined spheromak---a configuration called slow formation~\cite{jarboe83}. As this paper will discuss, the turbulence nature of this plasma is unaffected by the source physics as the selective decay of the plasma likely occurs locally.

Magnetic fluctuations are measured using an arrayed $\dot{B}$ probe; $\dot{B}$ in three directions $(r,\theta,z)$ is measured using 3 orthogonally-oriented $0.3$cm single-loop coils at 16 locations separated by $0.4$cm, beginning $1$cm from the cylindrical axis. Signals are acquired using a DTaq digitizer at 14-bit resolution and $65$MHz sampling rate. Magnetic field vectors are computed through numerical integration of the $\dot{B}$ signal. Mach number fluctuations, as a proxy for velocity fluctuations, are measured using a Mach probe oriented along the z-axis~\cite{Zhang11} and located $1.5$cm from the inner edge of the flux conserver. Density measurements are made using a He-Ne ($633$nm) laser interferometer through a diameter of the flux conserver set an axial distance of $21.5$cm from the midplane in the direction of the plasma source. In the data presented here, it is assumed that the plasma has already undergone initial Taylor relaxation by the time it reaches the midplane where most of the diagnostics are located. The chamber is pumped down to $\sim 6 \times 10^{-8}$Torr and undergoes a helium glow discharge between runs, so it is assumed that any influence of impurities on the plasma is low.

An additional important point regarding the SSX MHD wind tunnel experiment is that due to the nature of the plasma production method, the plasma does not have a dominant guide field direction; that is, the turbulence tends to be isotropic. This is evident in the similarity of the spectra for the three different probe directions, as discussed below.

\section{HiFi Simulation}

Simulations of the SSX MHD wind tunnel experiment have been performed with the HiFi spectral-element multi-fluid modeling framework \cite{Glasser04,Lukin08}.  Within HiFi, the following set of normalized compressible Hall-MHD equations with wall recycling is advanced in time:
\begin{eqnarray}
  \ptt{[\ln(\rho)]} &+& \frac{1}{\rho}\nabla\cdot
  \left\{\rho\vvec_i - D_\rho\rho\nabla[\ln(\rho)]\right\} = 0 
  \label{eq:Continuity}\\
  \ptt{\left(\rho\vvec_i\right)} &+&
  \nabla\cdot\left[\rho\vvec_i\vvec_i + (p_i + p_e)\mathbf{\bar{I}}
    - \mathbf{\Pi}_i -  \mathbf{\Pi}_e - D_\rho\nabla(\rho\vvec_i)\right] \nonumber \\ 
  &=& \Jvec\times\Bvec - R_w(\mathbf{r})\rho\vvec_i
  \label{eq:Momentum}\\
  \ptt{\Avec} &+& \frac{d_i}{\rho}\nabla\cdot\mathbf{\Pi}_e = \vvec_e\times\Bvec + \frac{d_i}{\rho}\nabla p_e - \eta\Jvec 
  \label{eq:OhmsLaw} \\
 \frac{\rho}{\gamma-1}\ptt{T} &+& \rho\nabla\cdot\left[\frac{1}{2}(\vvec_i + \vvec_e)T\right] 
    -  \nabla\cdot\mathbf{q} - \frac{1}{\rho}\nabla\cdot\left[\frac{D_\rho\rho^2}{\gamma-1}\nabla T\right]  \nonumber \\
    &=& \frac{1}{(\gamma-1)}\left[\frac{\gamma-2}{2}(\rho\vvec_i + \rho\vvec_e)\cdot\nabla T - R_w(\mathbf{r}) \rho T\right] \nonumber \\
    &+& \frac{1}{2}\left[\left(\mathbf{\Pi}_i + D_\rho\rho\nabla\vvec_i\right):\nabla\vvec_i + \mathbf{\Pi}_e:\nabla\vvec_e + \eta|\Jvec|^2 \right]
    \label{eq:Tempr}
 \end{eqnarray}
where
\begin{equation*}
  \Bvec = \nabla\times\Avec, \hspace{3mm} \Jvec = \nabla\times\Bvec, \hspace{3mm} \vvec_e = \vvec_i - (d_i/\rho)\Jvec, \hspace{3mm} p_i = p_e = \rho T,
\end{equation*}
$\rho$ is the plasma number density, $\vvec_i$ and $\vvec_e$ are the ion and electron velocity fields, $\Avec$ is the electro-magnetic vector potential, $\Bvec$ is the magnetic field, $\Jvec$ is the current density, $p_i$ and $p_e$ are the ion and electron pressures which are set equal to each other under the assumption of fast electron-ion temperature equilibration, $T$ is the temperature of either species,  $\mathbf{\Pi}_i=\mu_i\nabla\vvec_i$ is the ion viscosity tensor, $\mathbf{\Pi}_e=\mu_e\rho\nabla\vvec_e$ is the anomalous electron viscosity tensor, $\mathbf{q}=\kappa\nabla T$ is the thermal conduction heat flux, and $\gamma=5/3$ is the ideal gas adiabatic index.  These equations have been normalized in the standard fashion \cite{Lukin11} by choosing three unit normalization quantities, which are set to be the unit length $R_0=R=7.8$cm, the unit magnetic field strength $B_0 = 5$kG, and the unit number density $n_0=10^{16}$ per cubic cm, which also sets the normalized ion inertial length coefficient $d_i=(c/\omega_{pi0})/R_0 = 2.9\times 10^{-2}$.  This value for $d_i$ is used in the Hall-MHD simulations, while in the resistive-MHD simulations all two-fluid term are neglected by setting $d_i=0$.

The resistivity $\eta=\eta(\rho,T,|\Jvec|)$ used in the simulations is a combination of the classical Spitzer resistivity and the semi-empirical Chodura resistivity designed to capture anomalous electron drag at low plasma density and high current density \cite{Chodura75,Sgro76,Meier11}:
\begin{equation}
\eta(\rho,T,|\Jvec|) = \frac{\eta_0^{Sp}}{T^{3/2}} + \frac{\eta_0^{Ch}}{\sqrt{\rho}}\left[1 - \exp\left(-\frac{v_0^{Ch}|\Jvec|}{3\rho\sqrt{\gamma T}}\right)\right],
\end{equation}
with the normalization constants $\eta_0^{Sp} = 10^{-4}\Lambda\sqrt{2 m_p}\mu_0 e^{3/2}\frac{n_0^2}{R_0 B_0^4} = 2.7\times 10^{-4}$ for the Spitzer resistivity, $\eta_0^{Ch} = \frac{0.1 m_e}{e\sqrt{\epsilon_0 \mu_0}}\frac{1}{R_0 B_0} = 4.4\times10^{-3}$ and $v_0^{Ch} = \frac{\sqrt{m_p}}{e\sqrt{\mu_0}}\frac{1}{R_0\sqrt{n_0}}=2.9\times10^{-2}$ for the Chodura resistivity, expressed in the standard notation and with the Coulomb logarithm set to $\Lambda=10$.
 
The wall recycling terms proportional to $R_w(\mathbf{r})$ in Eq.~(\ref{eq:Momentum}) and Eq.~(\ref{eq:Tempr}) represent the loss of plasma momentum and temperature to the wall under the assumption of near-wall particles impacting the wall at some velocity but sputtering off the wall with negligible velocity, followed by instantaneous thermalization.  It is further assumed that all particles impacting the wall return into the plasma, an assumption that may be relaxed in future work.  For the simulations described here, $R_w(\mathbf{r})$ is given by:
\begin{equation} 
R_w(r,z) = \frac{1}{\tau_w}\left\{\exp\left[-\frac{z}{\lambda_w}\right] + \exp\left[-\frac{L/R_0 - z}{\lambda_w}\right] + \exp\left[-\frac{1-r}{\lambda_w}\right]\right\},
\end{equation}
where $\lambda_w= 0.05$, equivalent to 0.39~cm, is the characteristic thickness of the near-wall recycling layer in the cylindrical simulation domain given by $(r,z)\in [0,1]\times[0,L/R_0]$ with $L/R_0=11.1$.  Due to lack of a good theoretical estimate for the characteristic wall recycling time $\tau_w$ in the SSX experiment, a series of simulations systematically varying $\tau_w$ was conducted to empirically determine its value by comparing the magnetic field decay times in the simulations to those determined from the experimental magnetic field measurements discussed below.  (Note that since the wall recycling directly impacts the temperature of the plasma, it also significantly impacts the magnetic field resistive decay time via the temperature dependence of the Spitzer resistivity.)  As a result of this single-parameter study with all other input parameters held fixed, the wall recycling time has been set to $\tau_w=2$, equivalent to $1.4 \mu$s. 

Equations (\ref{eq:Continuity}-\ref{eq:Tempr}) also contain several terms proportional to a density diffusion coefficient $D_\rho$.  The specific form for these has been derived by taking velocity moments of a modified Boltzmann transport equation for the ion distribution function $f_i(t,\mathbf{r},\vvec)$ \cite{Krall86}; with the Boltzmann equation modified by adding an extra $D_\rho\nabla_{\mathbf{r}}^2 f_i$ diffusion term designed to represent sub-grid scale plasma mixing.

The simulations are performed on a cylindrical mesh with the spectral elements uniformly distributed in each direction with a grid of $(n_r,n_\theta,n_z)=(24,24,180)$ elements and the $3^{rd}$ order modified Jacobi polynomials used to expand the solution in each direction within each element \cite{Lukin08}.  Thus, the simulations' effective mesh size is $(N_r,N_\theta,N_z)=(72,72,540)$, providing radial resolution of 0.11~cm, angular resolution of 5 degrees, and axial resolution of 0.16~cm.

The simulations are initialized with the plasma at rest but far from an equilibrium, with nearly all of the plasma density, thermal energy, and magnetic field concentrated in one end of the tunnel, within $0 < z < 1.5$. This is intended to mimic the initial formation and the rapid expansion of the magnetized plasma cloud into the near-vacuum chamber in the SSX experiment. Specifically, the plasma number density is initialized with the following spatial distribution:
\begin{equation}
\rho(z)|_{t=0} = \left\{\begin{array}{l} 
1, \hspace{3mm} 0 \le z < (1 - \lambda_\rho) \\
\frac{1+\rho_{min}}{2} + \frac{1-\rho_{min}}{2}\sin\left(\frac{1-z}{2\lambda_\rho}\pi\right), \hspace{3mm} (1 - \lambda_\rho) \le z \le (1 + \lambda_\rho) \\
\rho_{min}, \hspace{3mm} (1 + \lambda_\rho) < z \le 11.1
\end{array}\right.
\end{equation}
where $\lambda_\rho=0.4$ is the half-width of the transition from the high-density region to the much lower density background tunnel plasma at $\rho_{min} = 10^{-2}$. The initial temperature distribution assumes an adiabatically expanded plasma with 
\begin{equation}
T(z)|_{t=0} = T_0\rho^{\gamma-1}(z)|_{t=0}
\end{equation}
and $T_0 = 0.1$, corresponding to the temperature of 6.2~eV in the high plasma density region and 0.3~eV in the rest of the tunnel. It should be noted that an immediate consequence of such a low temperature in the tunnel is a very high resistivity of that background plasma, which, in turn, is equivalent to the near-vacuum condition for the magnetic fields that begin to rapidly expand into the tunnel immediately upon release.

The initial condition for the vector potential $\Avec_0(\mathbf{r})$ prescribing the desired initial magnetic field configuration is generated by following the general prescription of \cite{Lukin11} and numerically solving within the HiFi framework $\nabla^2\Avec_0 = -\Jvec_0$ for $\Avec_0(\mathbf{r})$ given some current density source function $\Jvec_0(\mathbf{r})$ and subject to the boundary conditions $\hat{n}\times\Avec_0=\mathbf{0}$ and
$\nabla\cdot\Avec_0=0$, where $\hat{n}$ is the unit normal vector at the boundary surfaces.  The following functional form is chosen for $\Jvec_0$:
\begin{eqnarray}
\Jvec_0(\mathbf{r}) &=& S(\mathbf{r})
\lambda_{sph}\left[-\pi J_1(\alpha r)\cos(\pi z)\hat{r} \right.\nonumber\\
  &+& \left.\lambda_{sph} J_1(\alpha r)\sin(\pi z)\hat{\theta} 
  + \alpha J_0(\alpha r)\sin(\pi z)\hat{z}\right],
\end{eqnarray}
where $\lambda_{sph} = \sqrt{\alpha^2 + \pi^2}$,
$\alpha = 3.8317$ is the first zero of the $J_1$ Bessel function, and
\begin{eqnarray*}
S(\mathbf{r}) &=& S_r(\mathbf{r})\times S_z(\mathbf{r}), \\
S_r(r) &\equiv& \left\{\begin{array}{l} 
1, \hspace{3mm} r < (1 - \lambda_J) \\
\frac{1}{2}\left[1 - \cos\left(\frac{1-r}{\lambda_J}\pi\right)\right], \hspace{3mm} (1 - \lambda_J) \le r \le 1 \end{array}\right. \\
S_z(z) &\equiv& \left\{\begin{array}{l} 
\frac{1}{2}\left[1 - \cos\left(\frac{z}{\lambda_J}\pi\right)\right], \hspace{3mm} 0 \le z \le \lambda_J \\
1, \hspace{3mm} \lambda_J < z < (1 - \lambda_J) \\
\frac{1}{2}\left[1 - \cos\left(\frac{1-z}{\lambda_J}\pi\right)\right], \hspace{3mm} (1 - \lambda_J) \le z \le 1 \\
0, \hspace{3mm} 1 < z \le 11.1
\end{array}\right.
\end{eqnarray*}
is the shape function that localizes the current density source function within the high-density plasma region to represent a single $1:1$ aspect ratio Bessel-function spheromak configuration modified so that $\Jvec_0$ drops to $0$ at its edges over the width $\lambda_J=0.1$.  To break axisymmetry, an additional small tilt-like perturbation is introduced into the vector potential at $t=0$ such that
\begin{equation}
\Avec(\mathbf{r})|_{t=0} = \Avec_0(\mathbf{r})\left[1 + \delta r\cos(\theta)\exp(-z^2)\right],
\end{equation}
with $\delta=0.1$.

All transport coefficient other than resistivity are set to constant and uniform values throughout the simulations with $\kappa=10^{-2}$, $\mu_i=10^{-2}$, $\mu_e=10^{-3}$, and $D_\rho=10^{-4}$.  Free-slip hard wall boundary conditions are imposed on the ion flow with $\hat{n}\cdot\nabla(\hat{n}\times\vvec_i)=\mathbf{0}$ and $\hat{n}\cdot\vvec_i=0$, the diffusive thermal and density transport across the boundary is limited by setting $\hat{n}\cdot\nabla T=\hat{n}\cdot\nabla\rho=0$, and the perfect conductor with no charge accumulation boundary condition on the electro-magnetic fields is set by requiring $\hat{n}\times\Avec=\hat{n}\times\Jvec=\mathbf{0}$ and $\nabla\cdot\Avec=\nabla\cdot\Jvec=0$.

\section{Results}

\begin{figure}[!htbp]
\centerline{
\includegraphics[width=8.5cm]{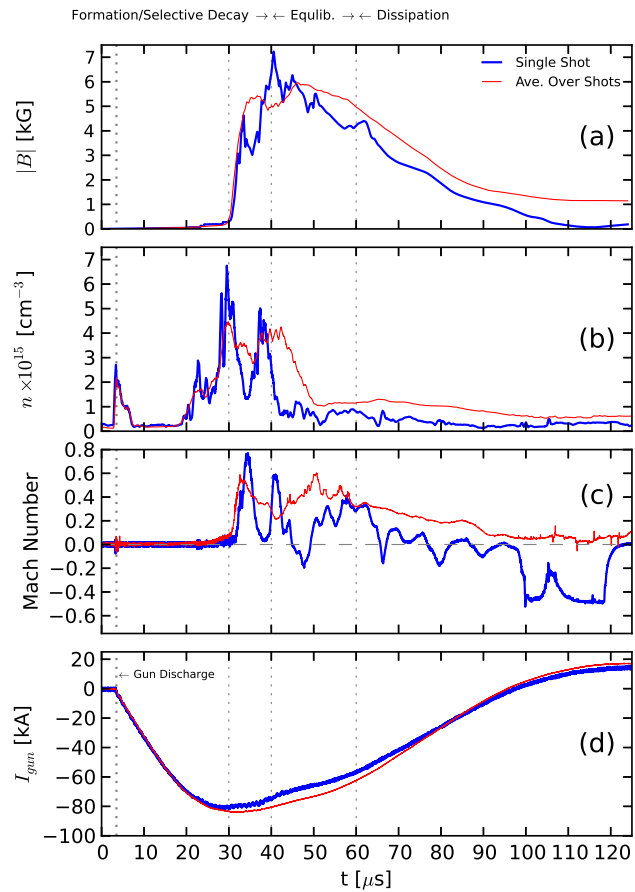}}
\caption{\label{fig:timeseries36} Time-series of (a) Magnetic Field magnitude, (b) Density, (c) Mach Number, and (d) Discharge current. An example single shot is shown (blue line) as well as the average trace for 35 shots (red line).}
\end{figure}

The SSX produces flux-rope plasma that persists on the order of $120\mu$s. Figure~\ref{fig:timeseries36} shows a time-series of numerically integrated B-field---$|B|=\sqrt{B_{r}^{2}+B_{\theta}^{2}+B_{z}^{2}}$---for the innermost tip ($1$cm from the cylindrical axis), interferometer line-averaged number density, $n$, Mach probe measurement of Mach number, $M$, and discharge current, $I_{gun}$, for a sample shot and for an average of 35 shots. For time analysis, the shots are divided into epochs to account for the dynamical nature of the plasma discharges as indicated by the dashed lines in Figure~\ref{fig:timeseries36}. The formation/selective-decay epoch spans from $30\mu$s to $40\mu$s. As shown by Gray, {\it et al.}~\cite{Gray13}, the selective-decay of the plasma into a Taylor state likely occurs in this time range. The equilibrium epoch ranges from $40\mu$s to $60\mu$s and is the period of turbulent fluctuations most closely analyzed here. This time range sees a generally stable period of fully developed turbulence balanced between the sourced plasma from the gun and resistive/viscous effects that degrade the magnetic fields and flows. The epoch from $60\mu$s on is considered the decay epoch and represents the flux-rope structure's resistive and viscous dissipation. Clearly, as in Figure~\ref{fig:timeseries36}(a), the magnetic field resistively decays away by $120\mu$s. Remaining unmagnetized plasma, however, has been shown to persist for many hundreds of microseconds.

\begin{table}
\caption{\label{tab:params}MHD wind tunnel plasma parameters during the equilibrium epoch for the present configuration of SSX.}
\begin{tabular}{|l|l|l|l|l|l|l|l|}
\hline
$V_{a}$&$f_{ci}$&Axial $\tau_{A}$&Radial $\tau_{A}$&$\rho_{i}$&$\delta_{i}$&$C_{s}$&$\beta$\\
\hline
$240km/s$&$7.6MHz$&3.5$\mu$s&0.3$\mu$s&0.091cm&0.51cm&31km/s&0.0967\\
\hline
\end{tabular}
\end{table}

In the equilibrium epoch, the average magnetic field is $5$kG and average density is $2\times 10^{15}$cm$^{-3}$; thus, radial and axial Alfv\'en transit times are $0.3\mu$s and $3.5\mu$s respectively. Ion temperature is measured using an ion Doppler spectrometer system at the midplane. Background ion temperatures (i.e. average temperatures made avoiding large heating events) are on the order of $20$eV. Though not measured in this dataset, previous measurements of the electron temperature suggest a value of $10$eV for this system. Combined with the average field and density values measured, estimates of ion gyrofrequency ($f_{ci}$), ion gyroradius ($\rho_{i}$) and ion inertial length ($\delta_{i}$) can be made. These values are listed in Table~\ref{tab:params}. Of particular note, the ratio of system size to $\rho_{i}$ is large, $R/\rho_{i} \cong 80$, suggesting that the influence of the wall on plasma dynamics is minimal.

The density trace in Figure~\ref{fig:timeseries36}(b) shows peaking in the formation epoch, but an eventual drop to approximately $10^{15}$cm$^{-3}$ for the majority of the discharge. The initial peaking is likely due to the initial spheromak formation, where plasma is being pushed into the tunnel, but before the threshold for break-off is achieved. The Mach number trace is on average positive, indicating flow predominately away from the gun source as would be expected for a single source configuration. The average Mach number during the equilibrium epoch is $M=0.4$ which for the sound speed $C_s$ listed in Table~\ref{tab:params} yields an average flow speed of $12$km/s. It is likely that flow speeds toward the center can be even higher. Indeed, a time of flight estimate based on the axial separation distance between the interferometer measurement and the midplane $\dot{B}$ measurement suggests a bulk plasma flow of $20$km/s. As seen in the single shot trace in Figure~\ref{fig:timeseries36}, there can sometimes be a flow reversal which likely indicates a reflection of flow off of the far axial boundary.


\subsection{Spectra}

\begin{figure}[!htbp]
\centerline{
\includegraphics[width=8.5cm]{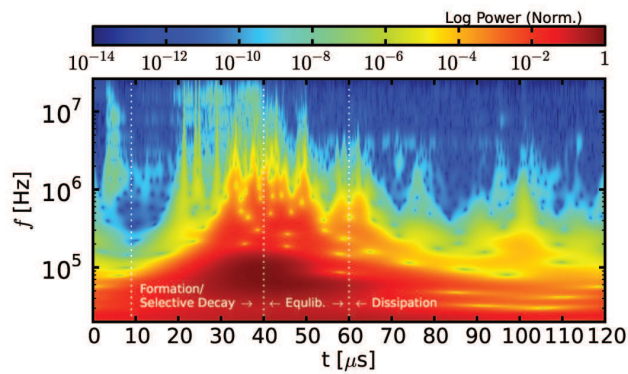}}
\caption{\label{fig:waveletcontour} Wavelet Power Spectrum of $\dot{B}_{\theta}$ as a function of time for a single shot (same as that shown in Figure~\ref{fig:timeseries36})}
\end{figure}

Given the dynamical nature of these plasmas, the spectral decomposition has been achieved using a Wavelet technique rather than with Fast Fourier Transforms (FFTs). This wavelet technique has been shown to be useful in situations where data may be non-stationary and provides a more accurate method for simultaneous spectral and temporal decomposition than a windowed Fourier transform because it applies a transform at many scales rather than a single scale~\cite{torrence98}. To achieve better resolution in frequency space, a sixth-order Morlet mother wavelet has been used. Morlet wavelet scales (i.e. frequencies) are generally well matched to Fourier scales, though it has also been shown that the choice of wavelet is not critical for determining power spectral densities (PSD)~\cite{torrence98}. A wavelet decomposition for a single shot $\dot{B}$ time-series is shown in Figure~\ref{fig:waveletcontour} with the wavelet scale converted to a Fourier frequency as the $y$-axis and time as the $x$-axis. The color scale corresponds to the normalized fluctuation power. Changes in the spectral nature can be seen with time: higher frequency fluctuations grow up and peak around $30\mu$s, hold stable until about $60\mu$s, and then begin to dissipate. This change in fluctuations supports the division of the data into epochs.

\begin{figure}[!htbp]
\centerline{
\includegraphics[width=8.5cm]{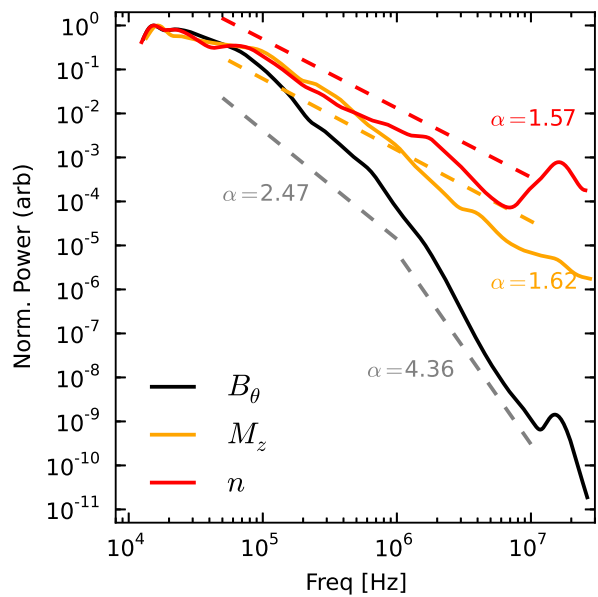}}
\caption{\label{fig:waveletspec} Wavelet Spectrum of B-field, density, and Mach number fluctuations for the equilibrium epoch, $40\mu$s to $60\mu$s.}
\end{figure}

The reduced power spectra for B-field, density and Mach number fluctuations are displayed in Figure~\ref{fig:waveletspec}. The B-field curves are produced by first taking a wavelet transform of the $\dot{B}$ time-series (for any one of the orthogonal components), yielding an array with $\dot{B}$ fluctuation power as a function of both time and frequency as in Figure~\ref{fig:waveletcontour}, and scaling the $\dot{B}$ power by $f^{2}$ in order to recover actual B-field fluctuation power. This approach is derived from the assumption that the B-field fluctuations can be Fourier decomposed such that
\begin{equation}
\frac{d}{dt}B(t) = \frac{d}{dt}B_{0}e^{i2\pi f} \sim ifB
\end{equation}
resulting in the relationship between $\tilde{B}(f)$ and $\tilde{\dot{B}}(f)$ as
\begin{equation}
|\tilde{B}(f)|^{2} = \frac{1}{f^{2}}|\tilde{\dot{B}}(f)|^{2}
\end{equation}
where $|\tilde{x}(f)|^{2}$ is the PSD. The same relationship applies for the Wavelet Transform. Once converted into a B-field power array, the 1D spectrum is calculated by summing over a certain time region - in this case the equilibrium epoch of $40\mu$s to $60\mu$s. The shot-averaged spectra for the magnetic field components ($B_{r}$, $B_{\theta}$, $B_{z}$) are nearly identical suggesting that magnetic fluctuations are isotropic. For clarity, only the $B_{\theta}$ component is shown in Figure~\ref{fig:waveletspec}.  A similar approach is taken for density and Mach number, though without the frequency scaling.

All three curves shown have roughly linear behavior in the spectral region between $50$kHz and $10$MHz. A power-law fit of spectral power to frequency ($P(f) \sim f^{-\alpha}$) is made for density and Mach number in this frequency range using a maximum likelihood estimation method (MLE)~\cite{clauset09} yielding exponents of $\alpha = 1.57$ for the density spectrum and $\alpha = 1.62$ for the Mach spectrum, which serves as a proxy for the velocity spectrum in this experiment. An exponent of $1.62$ for velocity is very close to the typical Kolmogorov exponent of $5/3 = 1.66$ for hydrodynamic flow fluctuations.

The magnetic field spectrum appears to have two separate regions: a shallower sloped region between $50$kHz and $1$MHz and a steeper sloped region between $1$MHz and $10$MHz. MLE fits for these regions give exponents of $\alpha = 2.47$ (shallower) and $\alpha = 4.36$ (steeper). Both the plots and the fits indicate that the spectra index for B-field fluctuations is steeper than that for Mach (velocity) fluctuations and thus steeper than that measured in the solar wind. This is perhaps indicative of a magnetic field dissipation mechanism that is present in the experimental plasma, but not in space.

The origin of the break in the spectra may come about in a number of ways. First, it may simply be due to the effect of flowing plasma across the probe. The Taylor hypothesis - frozen-in flow - suggests that measured temporal spectra can be mapped to spatial spectra when the medium containing the fluctuations is moving past a probe at some velocity. Thus, in the spectrum, there is some critical frequency below which fluctuations can be considered to be predominately due to spatial structure, and above which fluctuations are due to some combination of spatial and temporal structure. A rough estimate for this critical frequency can be made by considering the average flow and probe size. The average axial flow for the equilibrium epoch as indicated by the Mach probe and the assumed $C_s$ is $12$km/s; with a $3$mm probe size, the critical frequency would be $4$MHz. Thus, the break in the slope could be showing a transition between the power spectrum representing the energy transfer in both spatial and temporal scales (higher frequencies) and a spectrum due to the energy transfer in only spatial scales (lower frequencies), since lower frequency temporal fluctuations become less observable by the probe. 

\begin{figure}[!htbp]
\centerline{
\includegraphics[width=8.5cm]{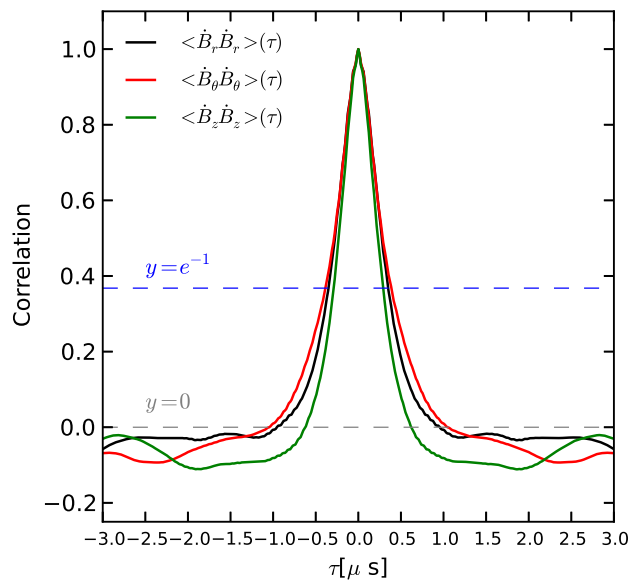}}
\caption{\label{fig:autocorr} Autocorrelation function for $\dot{B}$ fluctuations in the time range of $40\mu$s to $60\mu$s.}
\end{figure}

The break could also be related to the temporal decorrelation of the signal. Figure~\ref{fig:autocorr} shows the autocorrelation function for $\dot{B}$ for each of the three axes, again during the equilibrium epoch. A decorrelation time can be defined as the $\tau$ at which the autocorrelation function crosses zero. For $\dot{B}_{r}$ and $\dot{B}_{\theta}$, $\tau = 1.0\mu$s which corresponds to a frequency of $1$MHz - also called a decorrelation rate. If there are any wave modes in the plasma, they could influence the spectrum at frequencies below the decorrelation rate. Indeed, there has been some evidence that mode types can have an effect on energy transfer rates that manifest in the power spectrum~\cite{shaikh09}. Above the decorrelation rate, the fluctuations arise purely from the turbulence which may have a different energy transfer rate. Consequently, this may appear as a change in the spectral index around the decorrelation frequency.

Finally, this break may be due, of course, to a transition to the dissipation range of the spectrum. Possible dissipation scales include the ion gyroradius and ion inertial lengths which are $\sim 0.1$cm and $\sim 0.5$cm, respectively, for the equilibrium epoch. Dissipation could also occur directly in frequency space with coupling to the ion gyrofrequency, which here is $7.6$MHz. If the Taylor hypothesis is made, and frequencies are mapped to wavelengths using the average velocity, the break point occurs at a length of $1.2$cm; the ion inertial length scale maps to a frequency of $2.4$MHz.

\begin{figure}[!htbp]
\centerline{
\includegraphics[width=8.5cm]{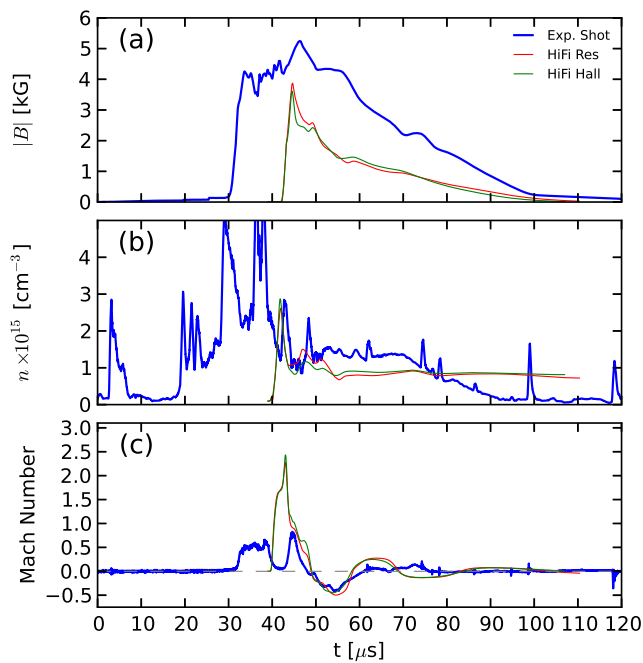}}
\caption{\label{fig:simtimeseries64} Time series of magnetic field, line-integrated density and Mach number for a resistive-MHD simulation, Hall-MHD simulation and a single experimental shot. In both the experiment and simulation, the magnetic and flow data is taken from the midplane while the line-integrated density is from a diameter 21.5cm from the midplane.}
\end{figure}

Synthetic diagnostics of the HiFi simulations of the SSX MHD wind tunnel also produce time series B-field, density and flow measurements, with the diagnostics location and probe sizes designed to resemble those used in the experiment as much as possible. For example, B-field at the mid-plane is measured by taking a volume average over a $0.3$cm diameter sphere corresponding to the size of the experimental probes; while the axial location of the chord-averaged plasma number density diagnostic, as well as the radial location of the mid-plane flow measurement, correspond to the locations of the respective experimental measurements.  However, unlike the experimental data, the simulated discharge consists of only a single spheromak that undergoes expansion, Taylor relaxation and decay with a lifetime on the order of $50\mu$s. The simulation does not presently model slow formation, or the sustainment of plasma injection seen in the experiment; however, a comparison between the two can be made by observing the similarity of the simulated pulse with the tail end of the experimental discharge. This can be clearly seen in Figure~\ref{fig:simtimeseries64} where the simulated time series has been shifted to approximately match the decay features of the B-field, as well as few fluctuation features of the density and Mach number. The experimental shot shown in Figure~\ref{fig:simtimeseries64} was chosen for its particularly well-matched features.

This similarity highlights the assertion that despite the longer injection time of plasma in the experiment (and thus, less well-defined large scale structure), the selective decay and self-organization still occurs. Moreover, the similarity allows us to compare the turbulent features of the experiment and the HiFi simulations.

\begin{figure}[!htbp]
\centerline{
\includegraphics[width=8.5cm]{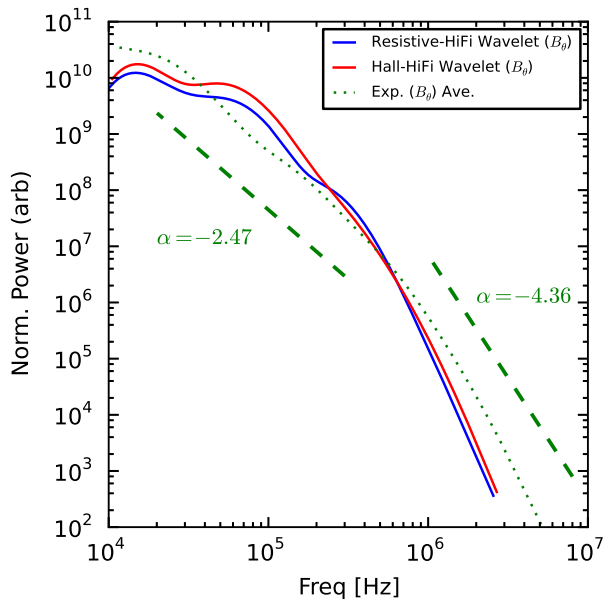}}
\caption{\label{fig:simBspectra} Comparison of B-field spectra from resistive-MHD and Hall-MHD simulations to the experimental spectra. The spectra are computed over a time period from $46\mu$s to $66\mu$s to approximate the equilibrium epoch of the shorter simulated pulses.}
\end{figure}

Figure~\ref{fig:simBspectra} shows wavelet spectra for the simulation runs as well as the averaged experimental spectra. Due to the smaller number of points in the simulation time-series, the wavelet decomposition used a fourth-order Paul mother wavelet, which has a slightly better time resolution than the sixth-order Morlet wavelet used above. The experimental data is also reanalyzed using the Paul wavelet for direct comparison to the simulation data. (Note that the differences between the two wavelet transforms is minimal for the experimental data, which is sampled at a higher frequency.) Like the experimental spectra, there appear to be two linear regions in the simulation spectra. The slopes of both the resistive-MHD and Hall-MHD simulation spectra are very close to that of the experimental data in the region between $20$kHz and $300$kHz. The simulation spectra, however, begin to steepen at a lower frequency than the experimental data: $\sim 300$kHz compared to $\sim 1$MHz. This is consistent with the autocorrelation function for the simulation which gives decorrelation times on the order of $3.3\mu$s. The simulations do not appear to shed any light on the possible role of dissipation. The spectra for both resistive-MHD and Hall-MHD simulations are similar, so that any modification of the underlying dynamics due to the inclusion of the Hall physics is not apparent in the data. Nevertheless, the slope of the simulation spectra are very close to the experiment.

\subsection{PDFs}

While spectra can be useful for obtaining the relative rate of energy transfer amongst injection, inertial and dissipation scales, they can obscure other signatures of turbulence that can only be seen when looking at higher integral moments. One technique for investigating these higher moments is to construct probability distribution functions (PDFs) from the data at different timescales and then to directly calculate the moments. The variation in the moments of PDFs as a function of timescale has been shown to be linear for fully-developed fluid turbulence~\cite{frisch95}. One can introduce a timescale into the time-series by filtering~\cite{frisch95,wan12} or by taking differences of the data points at different time steps~\cite{Greco08,Greco09}. This latter technique is used here, where a series of $\Delta\dot{B}$ increments is constructed for a variety of timesteps, $\tau$, as in
\begin{equation}
\Delta \dot{B}_{\tau}(t) = \dot{B}(t+\tau)-\dot{B}(t)
\label{eq:increments}
\end{equation}

\begin{figure}[!htbp]
\centerline{
\includegraphics[width=8.5cm]{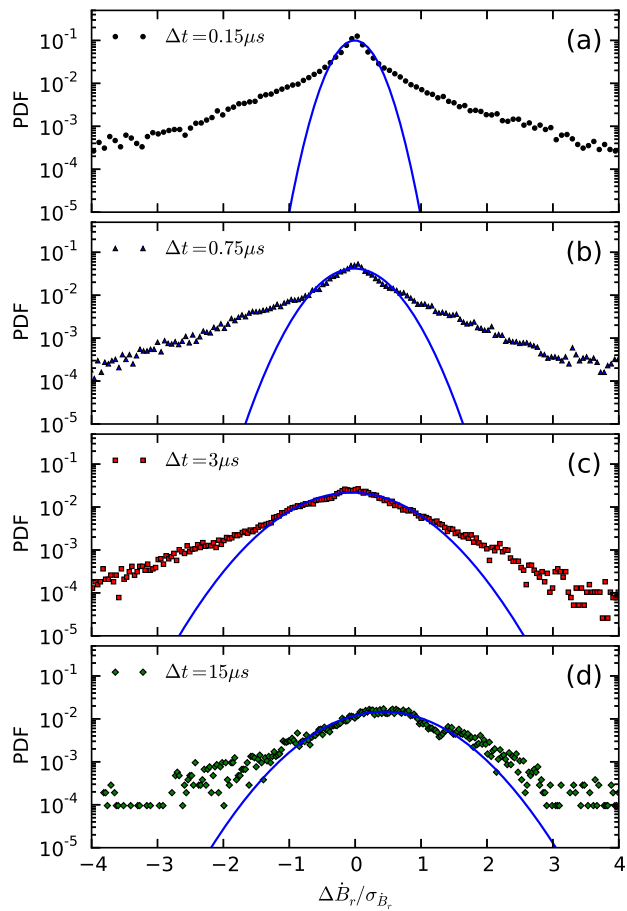}}
\caption{\label{fig:pdfs} PDFs of $\Delta \dot{B}$ in the time range 40 to 60 $\mu$s normalized to the standard deviation for each list and total integral of the histogram for different time increments: (a) $0.15\mu s$, (b) $0.75\mu s$, (c) $3.00\mu s$ and (d) $15.0\mu s$.}
\end{figure}
\begin{figure}[!htbp]
\centerline{
\includegraphics[width=8.5cm]{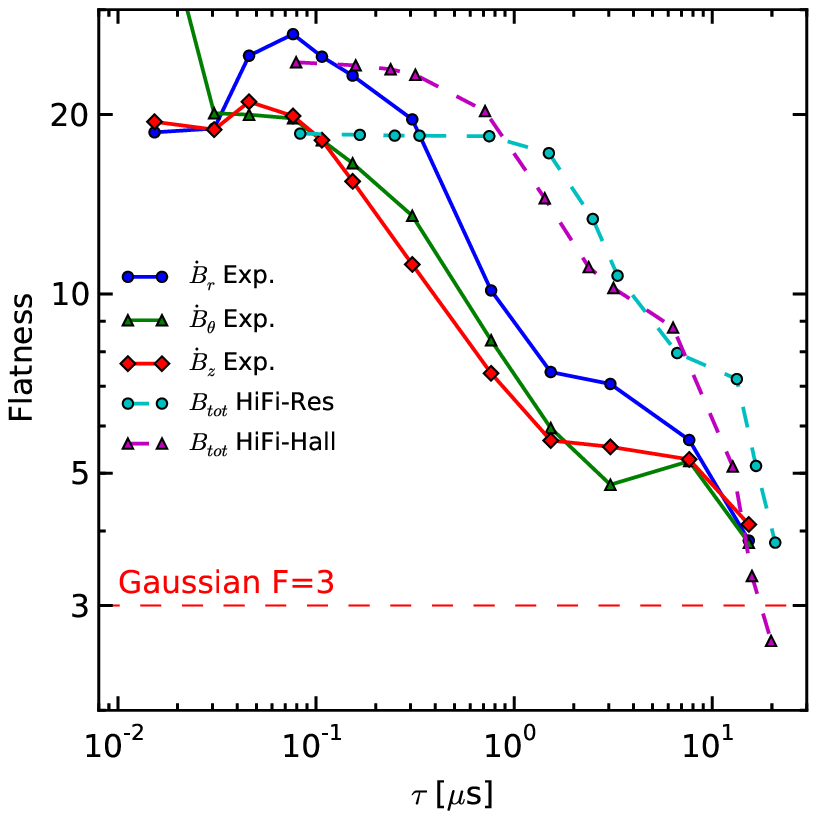}}
\caption{\label{fig:flatness} The flatness of the PDFs as a function of $\tau$ on a log-log scale. A flatness of $3$ would indicate a perfectly Gaussian distribution. A linear slope between $0.1\mu$s and $1\mu$s - corresponding to the frequency range of $1$MHz to $10$MHz - suggests a power-law like relationship between the flatness and the timestep. The flatness curves for the HiFi simulations are also listed. They show a similar shape, but offset in $\tau$.}
\end{figure}

A series of these PDFs for the equilibrium epoch time-series is shown in Figure~\ref{fig:pdfs} for $\tau$'s of $0.15\mu$s, $0.75\mu$s, $3\mu$s and $15\mu$s which corresponds to filtering the time-series above frequencies of $6.6$MHz, $1.3$MHz, $333$kHz, and $66$kHz respectively. The PDFs are normalized to the root mean square value of the increments and to the total integral of the histogram. A Gaussian function is fit to each distribution. One can clearly see a transition from non-Gaussian distributions for small $\tau$'s to more Gaussian distribution for larger $\tau$'s. The features of the non-Gaussian distributions are indicative of intermittency in the signal; the fat tails signify large swings in the values of the time-series and the timestep indicates the relative temporal size of the intermittent events.

The level of intermittency as a function of $\tau$ can be quantified using a calculation of the flatness defined as~\cite{deWit13}
\begin{equation}
F(\tau) = \frac{S_{4}(\tau)}{S_{2}(\tau)^{2}}
\label{eq:flatness}
\end{equation}
where $S_{p}(\tau)$ is the $p^{th}$ order moment for the distribution of increments, $\Delta \dot{B}$, as in
\begin{equation}
S_{p}(\tau) = \int_{-x_{max}}^{x_{max}} P(\Delta \dot{B})(\Delta \dot{B})^{p} dx
\label{eq:structurefunc}
\end{equation}
with $x$ representing a histogram bin width. An exact Gaussian distribution would have a flatness equal to $3$.

The resulting function for flatness is shown in Figure~\ref{fig:flatness} for each of the three $\dot{B}$ measurements. The plot shows the trend toward Gaussian (as represented by the dashed line at $F=3$) as the timestep is increased. An increase in flatness signifies an increase in the intermittency observed in the PDF - the flatness grows rapidly beyond $\tau = 1\mu$s corresponding to a frequency of $1$MHz. Since fluctuations beyond this frequency are more and more temporally correlated (as indicated by the autocorrelation function), it is clear that the intermittent nature is truly embedded in the magnetic signal. Conversely, a trend toward a Gaussian distribution might be expected for fluctuations that are not temporally correlated; however, recent work in this vein on solar wind measurements has speculated that such a change in flatness versus timescale might be expected at the transition from inertial to dissipation range turbulence~\cite{wan12}. It is also worth noting that the transition occurs around the autocorrelation time measured above. This is perhaps not unlike the spatial PDF measurement conducted in Greco, {\it et al}~\cite{Greco08} which showed a tendency toward Gaussian distributions beyond the radial correlation length whose analogue here would be the autocorrelation time.

The simulation data can also be analyzed within the PDF of increments framework. The flatness curve is also shown in Figure~\ref{fig:flatness} for the two different simulation runs. The shape of the curve is very similar to the experimental one, though the changes in flatness are offset in time increment as higher values of flatness are reached at larger increments. The largest values of flatness for both the experiment and the simulation are similar, suggesting that the mechanism responsible for the intermittency might be common to both.

\section{Conclusion}

We have hypothesized that the turbulent relaxation of a twisted flux rope plasma in the SSX wind tunnel was promoted by intermittency and coherent structures.  The rapidity of the relaxation process (about one Alf\'ven time) could be due to fast local relaxation of the coherent structures followed by a slower relaxation of the structure as a whole.  In this study, we have focused on time series and temporal frequency statistics: autocorrelation function, power frequency spectra, and PDFs of temporal increments.  Spatial correlations and spectra will be reported in a later article.

Magnetic field power frequency spectra from the experiment show power-law behavior over two decades but with steeper spectral indices than measured in the solar wind and predicted from theory. Comparisons of power frequency spectra with numerical simulations show good correspondence with indices $\alpha \approx 2.5$ at low frequencies and steeper at higher frequencies.  The reason for the break in the spectra slope has not been identified; however, both experimental and simulation data had measured decorrelation rates which could be placed near the respective break points. Furthermore, it is not clear whether a dissipation scale has been observed in the SSX plasma in these experiments.

Probability distribution functions of the magnetic field increments expose non-Gaussian higher order statistics connected to intermittency and coherent structures.  We find large values of flatness at short time increments corresponding to small spatial scales in both the experiment and the simulations.  In addition, we find that the shape of the curve of flatness with increasing time increment is also very similar in both the experiment and the simulations.

The observation of non-Gaussian values for flatness in the PDFs, and the trend for increasing flatness with decrease in timestep is a strong indication for the presence of intermittent fluctuations and/or coherent structures in the plasma, which could not be observed with spectra alone. The physical origin of these coherent structures in the SSX plasma has not yet been identified. As seen in previous simulation work~\cite{Servidio11b}, the coherent structures can be matched to the presence of current sheets which in turn could be the site of a reconnection layer. Given the observation of reconnection in past SSX work~\cite{Cothran09,Gray10}, there is a strong likelihood that such events are present in the current SSX configuration though perhaps at a smaller scale.

\section*{Acknowledgements}
  We gratefully acknowledge many useful discussions with William Matthaeus. This work has been funded by the US DoE Experimental Plasma Research program and the National Science Foundation.  The simulations were performed using the advanced computing resources (Cray XC30 Edison system) at the National Energy Research Scientific Computing Center.

\end{document}